%% file: 0_main.tex
\documentclass[runningheads]{llncs}
\usepackage[T1]{fontenc}
\usepackage{graphicx}
\graphicspath{{images/}}
\usepackage{hyperref}
\usepackage{soul}
\setstcolor{blue}
\usepackage{xcolor}
\usepackage{amsmath}
\usepackage{amssymb}
\usepackage[ruled,vlined,linesnumbered]{algorithm2e}
\usepackage{todonotes}
\usepackage{booktabs}

\newcommand{\mypar}[1]{\smallskip\noindent\textbf{#1.}}

\begin{document}
\title{The Impact of Event Data Partitioning on Privacy-aware Process Discovery}
\titlerunning{Impact of Event Data Partitioning on Privacy-aware Process Discovery}

    \author{Jungeun Lim\inst{1} \and
    Stephan A. Fahrenkrog-Petersen\inst{2,3} \and
    Xixi Lu\inst{4} \and
    Jan Mendling\inst{2,3} \and
    Minseok Song\inst{1}
    }
    \authorrunning{Lim et al.}
    \institute{Pohang University of Science and Technology, Pohang, South Korea \and
    Weizenbaum Institute, Berlin, Germany
    \and
    Humboldt-Universität zu Berlin, Berlin, Germany
    \and
    Utrecht University, Utrecht, The Netherlands
    \email{stephan.fahrenkrog-petersen@hu-berlin.de}
    }

\maketitle              %
\begin{abstract}
Information systems support the execution of business processes.
The event logs of these executions generally contain sensitive information about customers, patients, and employees.
The corresponding privacy challenges can be addressed by anonymizing the event logs while still retaining  \emph{utility} for process discovery. 
However, trading off utility and privacy is difficult: the higher the complexity of event log, the higher the loss of utility by anonymization.
In this work, we propose a pipeline that combines anonymization and \emph{event data partitioning}, where event abstraction is utilized for partitioning. By leveraging event abstraction, event logs can be segmented into multiple parts, allowing each sub-log to be anonymized separately. This pipeline preserves privacy while mitigating the loss of utility.
To validate our approach, we study the impact of event partitioning on two anonymization techniques using three real-world event logs and two process discovery techniques.
Our results demonstrate that event partitioning can bring improvements in process discovery utility for directly-follows-based anonymization techniques.

\end{abstract}

\input{1_intro}
\input{2_relwork}
\input{3_approach_stephan}

\input{4_evaluation_v2} %
\input{5_conclusion}

\bibliographystyle{splncs04}
\bibliography{6_refers}

\end{document}

%% file: 1_intro.tex
\section{Introduction}
\vspace{-0.3em}
Event logs recorded by information systems are the backbone of process mining. Typically, these event logs contain personal information~\cite{DBLP:conf/caise/VoigtFJKTMLW20} that need to be protected in terms of privacy~\cite{elkoumy2021privacy}. 
Such protection can be achieved through the anonymization of event logs~\cite{fahrenkrog2020pripel}. Anonymization aims to provide a formal privacy guarantee for the data and, at the same time, maximize the \emph{utility} of the anonymized event log. 
One way utility can be defined is by evaluating the quality of process models discovered using the anonymized log~\cite{rafiei2021towards}. 

Many anonymization techniques work through the insertion of noise~\cite{fahrenkrog2023semantics} with the aim of providing \emph{differential privacy}~\cite{dwork2006differential}. This noise is inserted into the result of queries that determine how often a control-flow behavior appears in the original event log. The query can either return a count of the trace-variants or the directly-follows relations within an event log~\cite{mannhardt2019privacy}.
A common goal of anonymization is to minimize the inserted noise while maximizing the preserved utility.
A high-level of utility is especially hard to reach for less-structured  processes. 

\begin{figure}[tb]
    \vspace{-1em}
    \centering
    \includegraphics[width=0.9\linewidth]{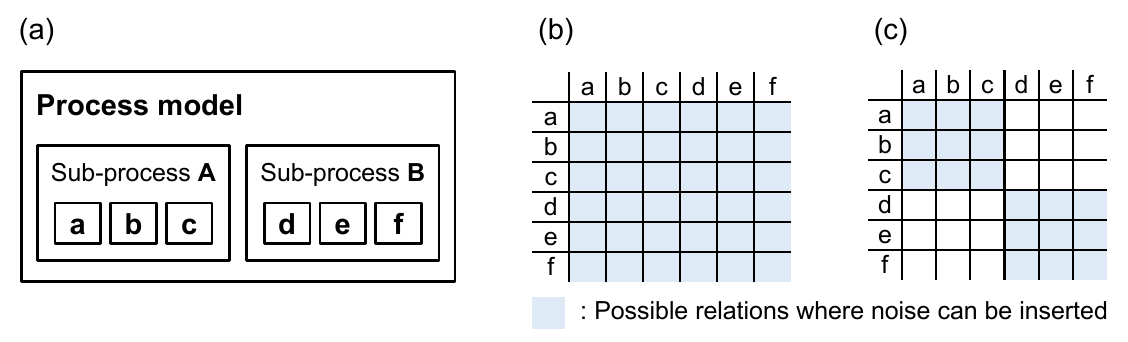}
    \caption{Noise insertion differences between considering and not considering process structure. (a): Structure of process model, (b): Noise insertion areas when not considering process structure, (c): Noise insertion areas when considering process structure.}
    \label{fig:intro}
    \vspace{-1.3em}
\end{figure}

In this paper, we build on the hypothesis that event data partitioning could facilitate a better trade-off between privacy and utility by reducing the amount of required noise. 
As an example, consider a process where sub-process $A$ consists of
activities $a$, $b$, and $c$, and sub-process $B$ consists of activities $d$, $e$, and $f$, as shown in \autoref{fig:intro} (a). If a privacy-preserving technique injects noise into the directly-follows relations without considering the process structure, the resulting noise can be extensive, as illustrated in \autoref{fig:intro} (b) in blue. 
However, when the structure is considered, as in \autoref{fig:intro} (c) in blue, the noise has only to be applied to each sub-process independently. The additional noise in \autoref{fig:intro} (b) could introduce unrealistic process behaviors, impacting process discovery utility.

In line with this argument, we propose a pipeline that first partitions an event log and then applies anonymization. The anonymized logs are subsequently used for process discovery. The utility of the anonymized log is assessed based on the quality of the discovered models.
First, we assess its effectiveness by measuring the utility of process models derived from sub-logs. 
We compare the utility of these models with those obtained from only anonymized event logs.
Second, we evaluate whether event data partitioning always provides utility benefits. To do so, we compare the utility of process models obtained by applying partitioning before anonymization versus those obtained by applying partitioning after anonymization.
We evaluated these questions for different types of anonymization, considering both directly-follows-based and trace-variant-based anonymization.
Our results show that performing event partitioning before anonymization leads to better process discovery utility for directly-follows-based anonymization. 

The remainder of this paper is structured as follows: \autoref{sec:relwork} reviews existing research on privacy-preserving process mining and event log abstraction. \autoref{sec:framework} introduces our proposed pipeline, and \autoref{sec:evaluation} presents the evaluation.
Finally, \autoref{sec:conclusion} summarizes the research and concludes the paper.

%% file: 2_relwork.tex
 \section{Related work}
\label{sec:relwork}
\vspace{-0.3em}
We outline the related work in this section. First, we give a short overview of privacy-preserving process mining in \autoref{sec:privacy} and outline related work in the area of event log abstraction in \autoref{sec:eventabs}.

\subsection{Privacy-aware process mining}
\label{sec:privacy}
Differential privacy~\cite{dwork2006differential} has been widely studied in the process mining literature~\cite{DBLP:journals/is/ElkoumyPD23,fahrenkrog2023semantics,DBLP:conf/rcis/RafieiWPA23}. 
One big advantage of differential privacy is that it is immune to post-processing. This means that process models generated from anonymized event logs are also secure against attacks like the one introduced by Kirchmann et al. \cite{DBLP:journals/corr/abs-2409-10986}.
The literature knows two types of mechanisms for control-flow anonymization: They anonymize either the \emph{directly-follows distribution} of an event log or the \emph{trace-variant distribution}. Most anonymization techniques are based on noise insertion. While many techniques exist for anonymization, the impact of pre-processing was not studied in the literature. 
The only exception is work by Elkoumy et al.~\cite{DBLP:conf/icpm/ElkoumyD22} that implemented an approach of privacy amplification. The idea is that through sub-sampling of traces a higher privacy guarantee can be given with less noise than if a differential privacy mechanism would be applied directly. However, this approach does not use process specific knowledge. Through our work, we present the first approach that utilizes event data partitioning in the pre-processing to achieve better utility for process discovery. 

A direction differing from the work on differential privacy is focused on group-based privacy guarantees. The main line of research here, focuses on generalization~\cite{DBLP:conf/caise/HildebrantFWR23}, substitution~\cite{DBLP:journals/dke/RafieiA21}, or the merging of events~\cite{DBLP:journals/dke/FahrenkrogPetersenAW23}. While we study our approach in combination with one differential privacy mechanism, it could also be applied in combination with one of these group-based privacy approaches. 

Furthermore, non-anonymization approaches for privacy have been explored. These are usually focused on deriving process models in settings where the event logs are distributed over multiple data owners. These approaches use techniques such as multi-party computation~\cite{DBLP:conf/caise/ElkoumyFDLPW20} and trusted execution environments~\cite{DBLP:conf/caise/GorettiBBC24}.
In contrast to these approaches, our technique can be applied to event logs stored by a single data owner.

\subsection{Event log abstraction}
\label{sec:eventabs}

Event log abstraction involves identifying sub-processes and their instances from low-level events \cite{lim2024navigating}. Various studies have proposed approaches for this, often addressing the identification of sub-processes and their instances separately. For identifying sub-processes, Baier et al. \cite{baier2014bridging} assumed that low-level activities with similar names originate from the same sub-process. Günther et al. \cite{gunther2010activity} assumed that low-level activities appearing close together in an event log likely originate from the same higher-level activity. Lu et al. \cite{lu2020discovering} proposed techniques that use domain knowledge and clustering.

Identifying sub-process instances has also been studied. A straightforward approach is to classify consecutive events with the same sub-process as originating from the same activity instance \cite{gunther2010activity,van2016enabling}. Other approaches include assuming that a sub-process can only be executed once \cite{lu2020discovering} or setting a time limit for the execution of a single instance to distinguish between instances \cite{baier2014bridging}.

In conclusion, the privacy-aware process mining literature did not focus on how pre-processing can be helpful for better utility. However, it was shown in the literature that pre-processing such as event log abstraction can lead to better process discovery utility. In this work, we address this gap in the literature by studying how event log abstraction impacts privacy-aware process discovery.

%% file: 3_approach_stephan.tex
\section{Anonymization utilizing Event Data Partitioning}
\label{sec:framework}

\begin{figure}
    \vspace{-1.5em}
    \centering
    \includegraphics[width=0.8\linewidth]{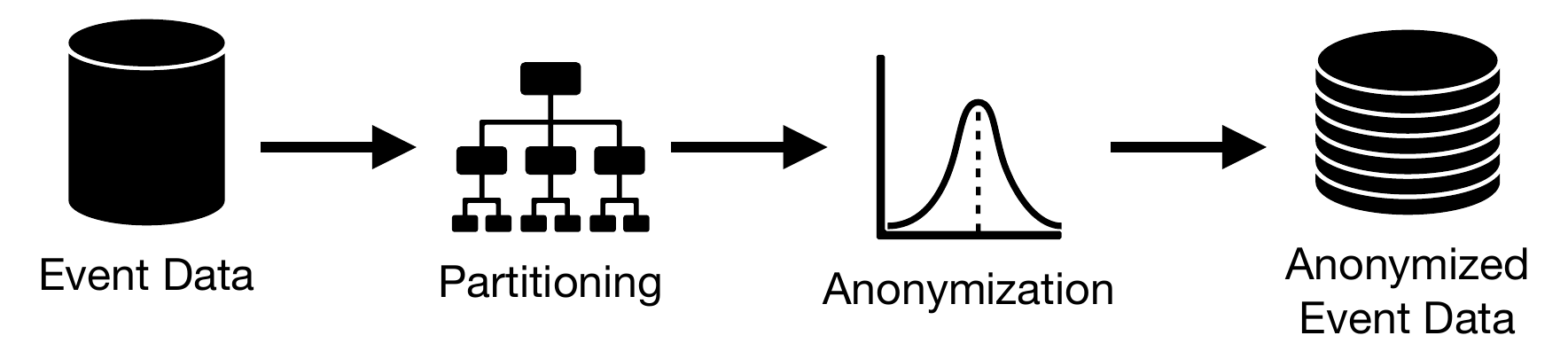}
    \caption{Anonymization pipeline.}
    \label{fig:framework_overview}
    \vspace{-2em}
\end{figure}

In \autoref{fig:framework_overview} we give an overview of the anonymization pipeline that allows us to combine event data partitioning and anonymization. In the remainder of this section, we elaborate on the different components of the pipeline.

\mypar{Event Data} Let us define an event as the execution of an activity $\alpha \in \mathcal{A}$, with $\mathcal{A}$ being the universe of all activities. 
Let a trace $\sigma \in \mathcal{A^*}$ be a sequence of events. Each trace represents the execution of a process. The sequence of a trace indicates in which order the events have been executed. We define an event log $L \in \mathcal{L}$ as a multi-set of traces.

\mypar{Partitioning} We turn to event data partitioning. First, we need to define event abstraction as a function $\psi: \mathcal{A} \mapsto \mathcal{A}$.
Let $\mathcal{A}$ always contain $\top$, which is the root activity, the highest-level activity within event abstraction. $\psi$ allows us to abstract low-level activities, such as the reading of a sensor, into higher-level activities. 
Note that multiple low-level activities can be mapped onto the same higher-level activity.
We assume that an activity $a$ can not be abstracted to itself $\psi(a) \neq a$, with the exception of $\psi(\top) = \top$.
The hierarchy $\psi$ can be either user-defined or derived through automated approaches such as clustering. We assume that when multiple low-level activities are abstracted into the same higher-level activity \( \alpha \), the higher-level activity \( \alpha \) represents a sub-process composed of these lower-level activities.
A hierarchy $\psi$ has to be defined such there are no cycles, e.g. $\psi(a) = \alpha$, $\psi(\alpha) = a$, and every activity can eventually be abstracted to the root activity $\top$, e.g.  $\psi(a) = \alpha$, $\psi(\alpha) = \top$.

\begin{figure}
    \centering
    \includegraphics[width=1\linewidth]{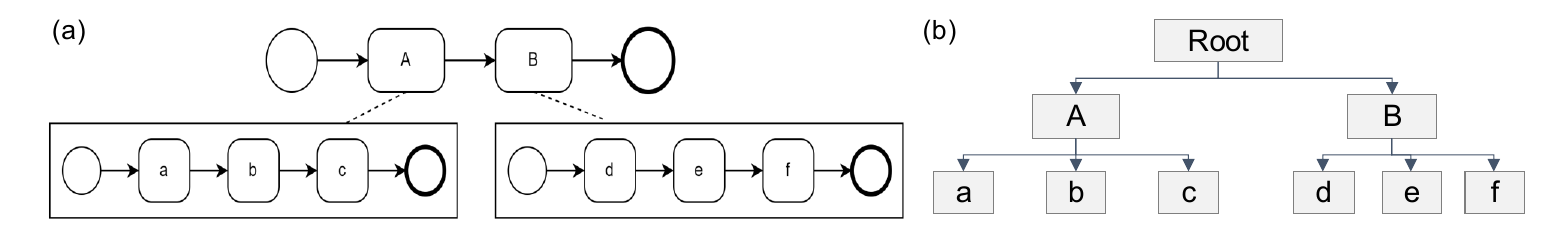}
    \caption{Running example for process with sub-processes.}
    \label{fig:framework_running_example}
    \vspace{-1em}
\end{figure}

 Our running example shown in \autoref{fig:framework_running_example}, illustrates such a process with such an event abstraction hierarchy. Here, the activities $\{ a,b,c\}$ can be abstracted to the activity $A$. 
 Now we could apply the function $\psi$ to every event within a trace $\sigma$ to create an abstracted trace $\sigma'$. In such a scenario, all information that can be derived from the low-level activities would be lost. Therefore, it would be better to \emph{decompose} a log $L$ into multiple sub-logs: $\{L_\psi,L_{\alpha_1},L_{\alpha_2}\}$ where $L_\psi$ is the event log consisting of the higher-level activities while $L_{\alpha_1}$ contains all the activities belonging to sub-process $\alpha_1$. These composed logs can be generated by the \emph{event data partitioning} function $\psi_L : \mathcal{L} \mapsto 2^{\mathcal{L}}$. 
 
 In \autoref{alg:abstraction}, we outline how $\psi_L$ works. First, we initialize the result set of sub-logs $S$ and the abstracted event log $L_\psi$ (see line~\ref{line:empty_result}-\ref{line:empty_result_log}). Next, we iterate over every trace $\sigma$ in our event log $L$ (see \autoref{line:iterate_trace}). Next, we create the set $C$ that is used to store traces of the sub-processes (see \autoref{line:empty_set_subprocess}). Now, we check for every activity of trace $\sigma$ if it can be abstracted using $\psi$ (see line~\ref{line:for_all_activities}-\ref{line:can_abstract}). If an activity $\sigma(i)$ can be abstracted, we abstract it within the trace $\sigma$ (see \autoref{line:get_abstract} \& \autoref{line:assign_abstract}) and save the sub-process information belonging to the sub-process $\alpha'$ in a corresponding trace $\sigma_\alpha'$ (see line~\ref{line:get_old_subprocess_trace}-\ref{line:update_old_subprocess_trace}). Such a procedure, leads to many occurrences of the sub-process activity $\alpha'$ in the trace $\sigma$. We apply a \emph{compression function} that removes every occurrence except the first and last of activity $\alpha'$ from the trace $\sigma$, signaling the start and end of the sub-process (see \autoref{line:compress}), while the actual information is encoded in the sub-process trace $\sigma_{\alpha'}$. We add the trace $\psi$ to the abstracted event log $L$ (see \autoref{line:add_xi}) and add the sub-process traces stored in the set $C$ to the set of event logs $S$ (see \autoref{line:update_C}).

\vspace{-1.5em}
\begin{algorithm}
    \SetAlgoLined
        \KwIn{An event log $L$} 
        \KwOut{Set of event logs $S\in2^{\mathcal{L}}$}
        $S\gets \emptyset$; \tcp{Defines the empty set of new event logs}
        \label{line:empty_result}
        
        $L_\psi\gets \emptyset$;
        \label{line:empty_result_log}

        \For{$\sigma \in L$}{
        \label{line:iterate_trace}
        
                $C \gets \emptyset$; 
                \label{line:empty_set_subprocess}
                
                \For{$i \in [1, \dots ,|\sigma|]$}{
                \label{line:for_all_activities}
                
                    \If{$\psi(\sigma(i)) \neq \emptyset$}{
                    \label{line:can_abstract}

                        $\alpha' \gets \psi(\sigma(i))$;
                        \label{line:get_abstract}

                        $\sigma_{\alpha'} \gets C(\alpha') \text{ if } \sigma_{\alpha'} \in C; \text{ otherwise } \sigma_{\alpha'} \gets \emptyset$;
                        \label{line:get_old_subprocess_trace}

                        $C \gets (C \setminus \sigma_\alpha) \cup \sigma_\alpha' \circ \sigma(i))$;
                        \label{line:update_old_subprocess_trace}

                        $\sigma(i) \gets \alpha';$
                        \label{line:assign_abstract}

                    }
                }
            $compress(\sigma)$;
            \label{line:compress}

            $L_\psi \gets L_\psi  \cup \sigma;$
            \label{line:add_xi}

            $S.update(C);$
            \label{line:update_C}
        }
        $S \gets S \cup L_\psi;$
        
        \Return $S$
    \caption{Event Log Partitioning function $\psi_L$}
    \label{alg:abstraction}
\end{algorithm}
\vspace{-1.5em}

\mypar{Anonymization}
Next, we anonymize the partitioned event data. We apply the notion of differential privacy. We define $\gamma: \mathcal{L} \mapsto \mathcal{L}$ as an anonymization function that transforms an event log into an anonymized event log. Further, 
let us define $img(\gamma) \subseteq \mathcal{L}$ as the 
\emph{image} of $\gamma$, i.e., the set of all event logs that may be returned 
by $\gamma$. Finally, we define two event logs $L_1, L_2 \in \mathcal{L}$ to be 
\emph{neighboring}, if they differ by exactly the data 
of one individual. In our setting, this corresponds to one trace,  i.e., $|L_1\setminus L_2| + |L_2\setminus L_1| = 1$. 
Based on~\cite{dwork2006differential}, we use the following definition for differential 
privacy:
\begin{definition}[Differential Privacy]
	\label{def:ldp}
	Given an anonymization function~$\gamma$ and privacy parameter $\epsilon 
	\in \mathbb{R}$, function $\gamma$ provides $\epsilon$-differential privacy, if for all neighboring pairs of event logs $L_1, L_2 
	\in \mathcal{L}$ and all subsets $\rho_\gamma \subseteq img(\alpha)$, 
  it holds 
	that:
\begin{equation*}
Pr[\gamma(L_1)\in \rho_\gamma] \leq e^\epsilon \times Pr[\gamma(L_2)\in 
\rho_\gamma] 
\end{equation*}
	where the probability is taken over the randomness introduced by the 
	anonymization function $\gamma$.
\end{definition}

We can apply any implementation of $\gamma$ that gives the differential privacy guarantee. An important rule of differential privacy is parallel composition. This rule states that different datasets that have no intersection can be anonymized independently with $\epsilon$-differential privacy their combination also is $\epsilon$-differential private. Therefore, we can anonymize the different event logs created by the event log partitioning independently. 

\mypar{Anonymized Event Data}
After anonymization, we end up with multiple anonymized event logs, i.e., $L'_\psi$, $L'_{A}$ and $L'_{B}$. The anonymized log $L'_\psi$ shows the process consisting of activities of higher level.  The other logs contain the data of the anonymized sub-processes. For each of these logs, we apply a discovery algorithm and evaluate the quality of the discovered process model to measure the \emph{utility}. Note that the other way around is also possible, namely, we first anonymize the event log and then partition the log into multiple sub logs. In the evaluation, we investigate the difference in the resulting utilities (RQ. 2).

%% file: 4_evaluation_v2.tex
\section{Evaluation}
\label{sec:evaluation}

In this section, we evaluate our proposed pipeline. Specifically, we examine the results for both directly-follows-based and trace-variant-based anonymization techniques. We focus specifically on the following two aspects:
\begin{description}
    \item[RQ1] How does performing partitioning before anonymization affect the utility of anonymized logs for process discovery?
    \item[RQ2] How does the order of partitioning and anonymization impact process discovery utility?
\end{description}

In \autoref{sec:exp_setup}, we outline our experimental design and the datasets we employ. %
In \autoref{sec:results}, we present the results of our experiments. Finally, in \autoref{sec:trade_offs}, we provide a qualitative discussion on the trade-offs of combining partitioning with anonymization.

\subsection{Experimental Setup}
\label{sec:exp_setup}
\begin{figure}
\vspace{-1em}
    \centering
\includegraphics[width=1\linewidth]{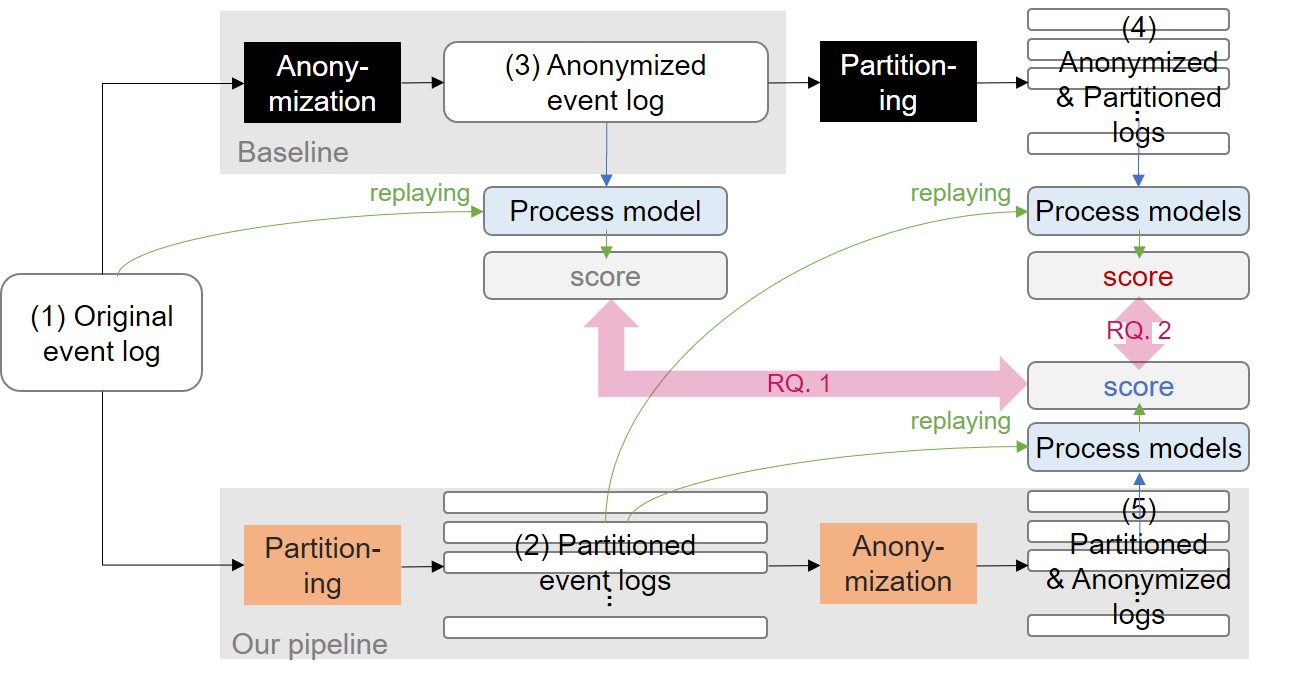}
    \caption{Experimental setup overview.}
    \label{fig:eval_overview}
    \vspace{-2em}
\end{figure}

We provide an overview of our experimental setup in \autoref{fig:eval_overview}. The first question (RQ1) aims to investigate the effect of our proposed pipeline (i.e., first partitioning and then anonymization, highlighted in orange in~\autoref{fig:eval_overview}) on utility (i.e., the quality of the process models discovered using anonymized logs). One may argue it is unfair to compare the utility of a set of logs to one log. Therefore, in the second question (RQ2), we examine whether any observed improvement is solely due to the introduction of partitioning rather than its specific placement in the pipeline. To investigate this, we compare the resulting utility of partitioning before (highlighted in orange) and after (highlighted in black) anonymization.

\mypar{Datasets}
For our evaluation, we used the BPIC2012~\cite{bpic12}, BPIC2015~\cite{augusto2018automated}\footnote{We used filtered log described in~\cite{augusto2018automated}.
Although five datasets are available, we only used the first one, BPIC\_1f.}, and BPIC2017~\cite{bpic17} event logs. 
All three event logs stem from real-world business processes. Both BPIC2012 and BPIC2017 describe a loan application process. BPIC2015 describes a permit application process used by Dutch municipalities.
All of these logs are known to consist of events from multiple sub-processes and have therefore been widely used in various event abstraction studies \cite{li2023event,rebmann2022gecco,lu2020discovering}. In \autoref{tab:datasets}, we provide descriptive statistics of these datasets.

\begin{table}
    \vspace{-1.2em}
    \centering
    \caption{Statistical description of the datasets.}\label{tab:datasets}
    \begin{tabular}{l@{\hspace{10pt}}r@{\hspace{10pt}}r@{\hspace{10pt}}r@{\hspace{10pt}}r@{\hspace{10pt}}r}
    \toprule
        Data & \#case & \#events & \#acts & avg. e/c & \#trace variants\\
    \midrule
        BPIC2012 & 13,087 & 262,200 & 24 & 20 & 4,366\\
        BPIC2015 & 902 & 21,656 & 70 & 24 & 295\\
        BPIC2017 & 31,509 & 1,202,267 & 26 & 38 & 15,930\\
    \bottomrule
    \end{tabular}
    \vspace{-1.2em}
\end{table}

\mypar{Applied techniques for event data partitioning}
During event data partitioning, we utilized three techniques for the higher-level activity mapping function ($\psi$ in \autoref{sec:framework}).
The first technique developed by Lu et al.~\cite{lu2020discovering} integrates domain knowledge into event abstraction (FHM, FlexHMiner). Second, we use a technique by G\"unther et al.~\cite{gunther2010activity} based on activity clustering (AC). Finally, we employ a random clustering that allows us to study if we also experience benefits with less sophisticated abstraction techniques.
We assumed that a high-level activity could only be executed once and distinguished between the start and end of higher-level activities, following the approach in~\cite{lu2020discovering}.

\mypar{Applied techniques for anonymization}
For anonymization, we used two anonymization techniques: one based on the Laplace mechanism (DF-Laplace)~\cite{mannhardt2019privacy} (since it does not require an additional parameter, as does the technique based on the exponential mechanism such as SaPa~\cite{fahrenkrog2023semantics}); the other one based on SaCoFa~\cite{fahrenkrog2023semantics} for the trace-variant query, since it provides the highest utility of all techniques that add additional behavior to anonymized logs. We used public implementations of both techniques~\cite{DBLP:conf/icpm/KirchmannFKAW22}. 
For differential privacy, we used the settings of $\varepsilon = \{ 0.01, 0.1, 1.0\}$ to test different orders of magnitudes for privacy guarantee. Lower values of $\varepsilon$ mean stronger privacy protection.
In our result figures, we plot the results of these settings within one bin, due to space limitations.

SaCoFa anonymizes trace-variants by constructing a prefix tree that accounts for the harmfulness of the prefix. The upper bound on the trace-variant length ($l$) was set to 50 across all settings. However, the privacy guarantee ($\varepsilon$) and the pruning parameter ($p$) were evaluated at three different combinations: (0.01, 300), (0.1, 200), and (1.0, 100) for BPIC2012 and BPIC2017, and (0.01, 100), (0.1, 100), and (1.0, 50) for BPIC2015.
To account for non-determinism of noise injection, we 
repeated the anonymization process 10 times for each setting.

\mypar{Applied techniques for process discovery}
For process discovery, we used Inductive Miner \cite{leemans2014discovering} and Heuristic Miner \cite{5949453}, setting the noise threshold to 0.2 for both techniques. Both techniques are widely used process discovery algorithms and we used their implementation in PM4Py~\cite{berti2019process}.

\mypar{Evaluation measures}
We evaluated the utility of the process models using four metrics. For fitness, we utilized token-based replay fitness~\cite{van2016data}. For precision, we employed ETC Precision~\cite{munoz2010fresh}. We also calculated the F1-score, which is the harmonic mean of fitness and precision. Finally, for generalization, we adapted the technique from \cite{augusto2018automated} with $k=3$.

When evaluating the multi-level process model, we followed the approach in \cite{lu2020discovering}. This involves calculating the performance for each sub-process in the multi-level process model and then averaging them.

\subsection{Results}
\label{sec:results}
\begin{figure} [htb]
    \vspace{-2em}
    \centering
    \includegraphics[width=0.9\linewidth]{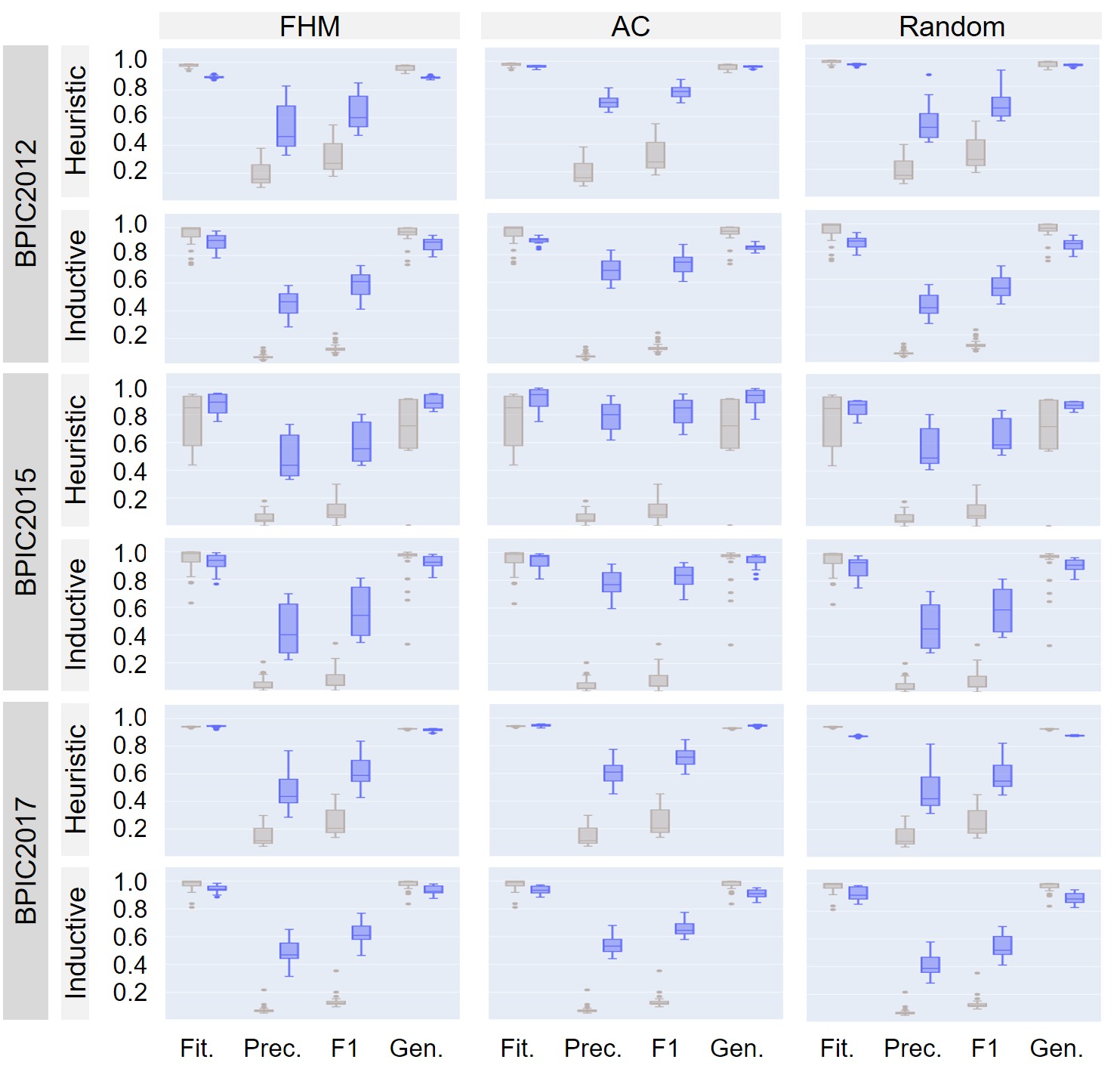}
    \caption{Utility comparison with DF-Laplace: Anonymization only (gray) vs. Partitioning and Anonymization (blue)}.
    \label{fig:df_eval35}
\end{figure}

\mypar{RQ1: Effect of Partitioning on DF-Laplace}
\autoref{fig:df_eval35} compares the utility of performing only anonymization with that of performing anonymization after partitioning, categorized by dataset, applied technique in event partitioning, and process discovery technique.
The fitness was higher when only anonymization was applied, for almost all datasets.
On the other hand, precision exhibited significant benefits from event partitioning: it was below 0.2 when only anonymization was performed, but increased to as high as 0.6 after partitioning at most. 
Overall, the increase in precision contributed to higher F1-score values when anonymization was preceded by event partitioning. 
Generalization showed good utility in most cases, with values exceeding 0.8 regardless of whether partitioning was applied, and differences remained within 0.2. 
These results indicate that incorporating partitioning, particularly with the DF-Laplace anonymization technique, can enhance precision while maintaining good levels of fitness and generalization.

\begin{figure} [htb]
    \centering
    \includegraphics[width=0.9\linewidth]{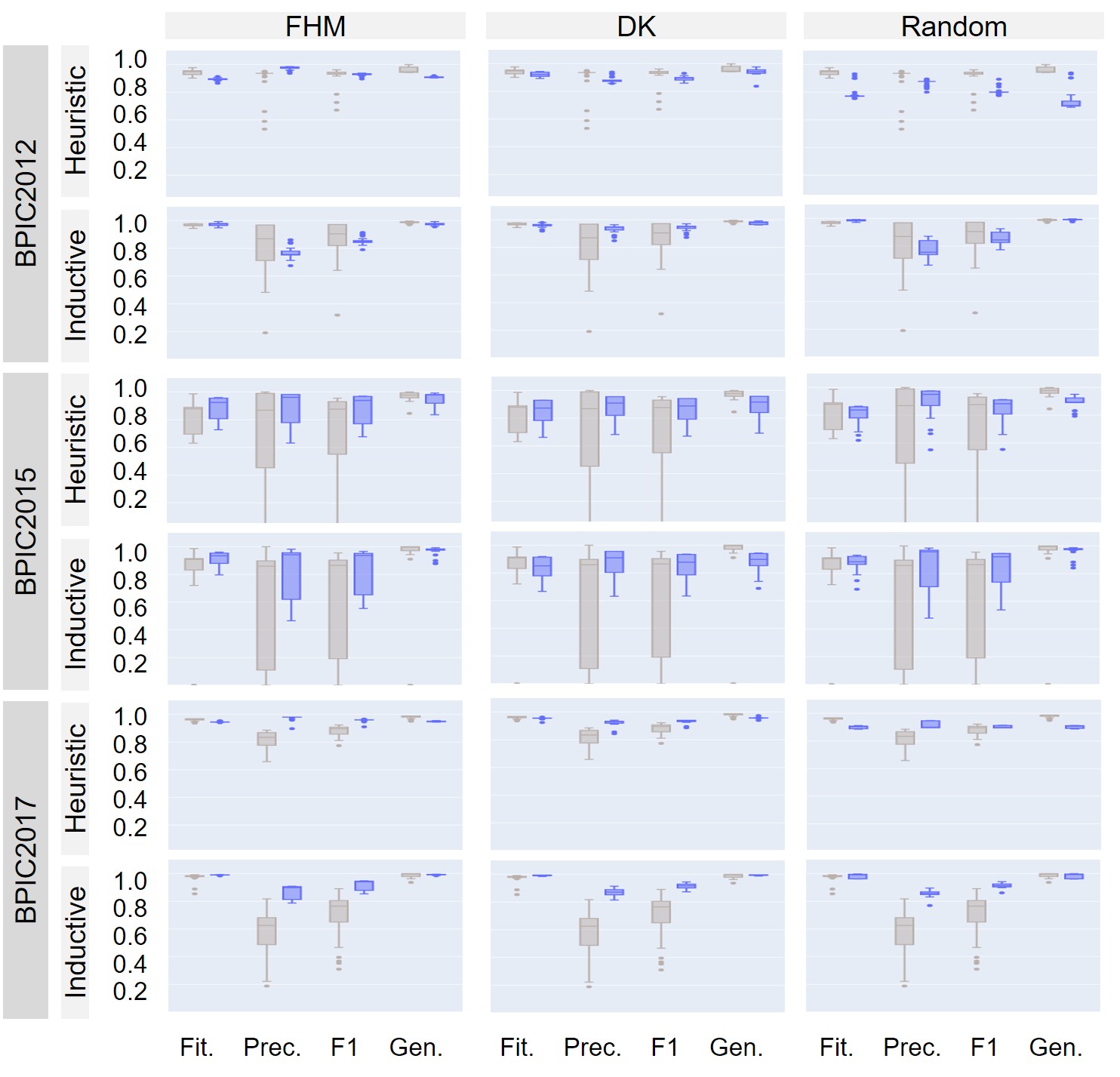}
    \caption{Utility comparison with SaCoFa: Anonymization only (gray) vs. Partitioning and Anonymization (blue).}
    \label{fig:sacofa_eval35}
    \vspace{-2em}
\end{figure}

\mypar{RQ1: Effect of Partitioning on SaCoFa}
When SaCoFa was used for anonymization, the results differed from those observed with DF-Laplace. While partitioning tended to improve the utility of anonymized logs when DF-Laplace was used, it appeared to have little to no effect on utility when SaCoFa was applied. As shown in \autoref{fig:sacofa_eval35}, there was generally little difference across all metrics between applying anonymization alone and applying partitioning before anonymization. Specifically, when DF-Laplace was used, precision showed a significant improvement after partitioning, but with SaCoFa, the difference in precision between logs with and without partitioning was minimal.

This outcome can be attributed to the fact that DF-Laplace significantly reduces precision when applied to the full log, whereas SaCoFa maintains relatively stable precision levels, and in some cases even improves them.

For the BPIC2017 dataset, however, precision showed an improvement compared to other datasets. This can be explained by the relatively low precision of the BPIC2017 log after anonymization, allowing partitioning to have a more noticeable effect.

For the BPIC2015 dataset, precision itself did not improve significantly, but partitioning effectively reduced its variation. This suggests that while anonymization alone results in good precision on average, there are cases where its quality is not consistently guaranteed. In such instances, partitioning can help stabilize the quality of anonymization and ensure more reliable results.

\begin{figure} [htb]
    \centering
    \includegraphics[width=0.9\linewidth]{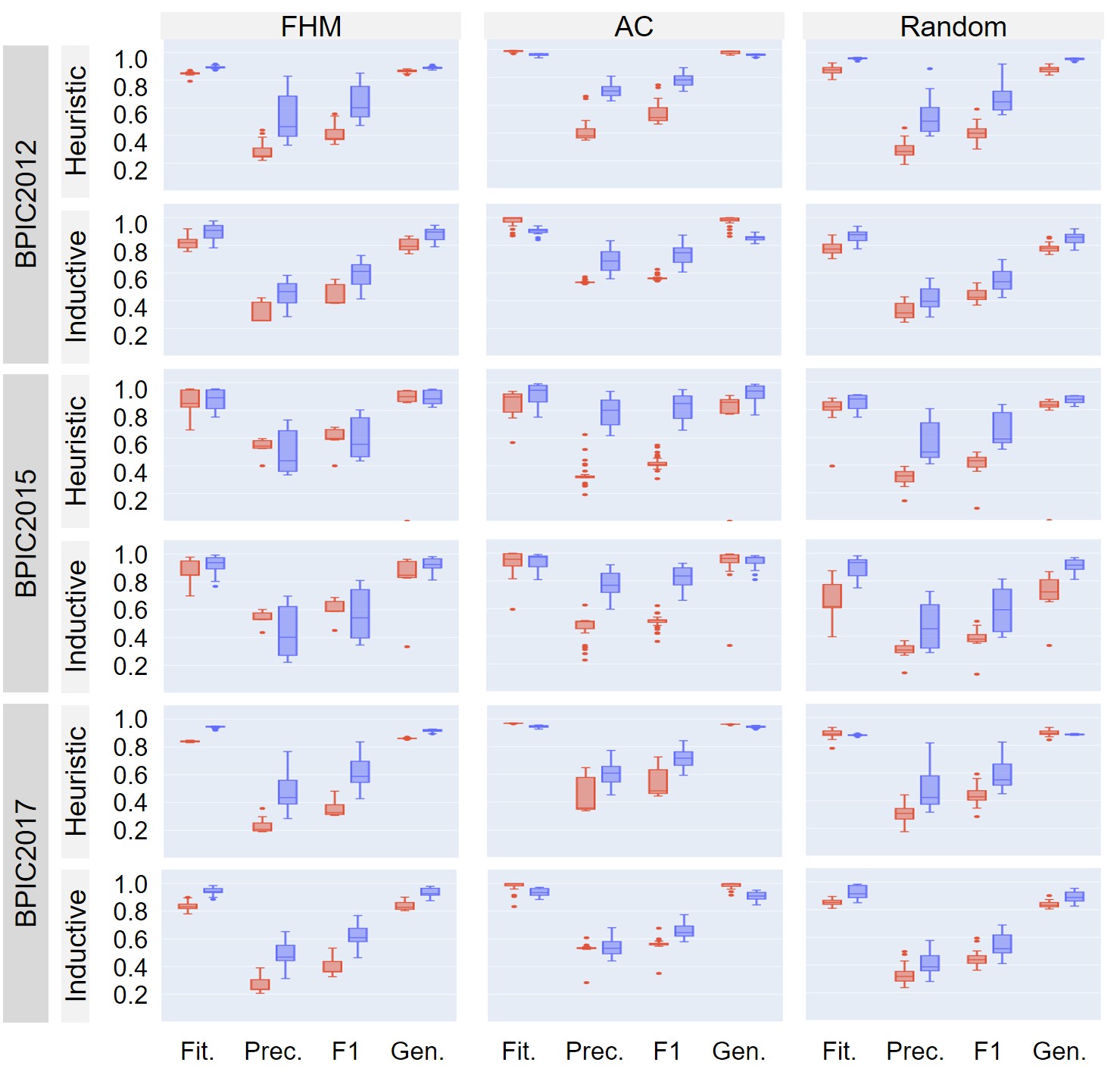}
    \caption{Utility comparison with DF-Laplace: Anonymization and Partitioning (red) vs. Partitioning and Anonymization (blue).}
    \label{fig:df_eval45}
    \vspace{-2em}
\end{figure}

\mypar{RQ2: Effect of Partitioning Order on DF-Laplace}
\autoref{fig:df_eval45} compares the utility of performing partitioning after anonymization and the utility of performing partitioning before anonymization, categorized by dataset, applied abstraction technique in event partitioning, and process discovery technique. Overall, across all datasets, performing partitioning before anonymization yielded better performance in all metrics compared to partitioning after anonymization. 
However, fitness and generalization generally exceeded 0.8 in both cases, resulting in relatively minor differences, while precision showed a more pronounced disparity. For example, when partitioning was performed using random clustering on BPIC2015 and the process model was discovered using the heuristic algorithm, the precision improved significantly—from approximately 0.3 when partitioning was performed after anonymization to 0.8 when partitioning was performed beforehand. In conclusion, these results demonstrate that when using DF-Laplace for anonymization, partitioning before anonymization helps to preserve more information in the data, thereby enhancing process discovery performance.

When partitioning was performed using FHM on BPIC2015, the results for precision differed somewhat from other observations. While precision generally improved significantly when partitioning was performed first, in this case it decreased. Considering that the precision was high when FHM was performed on BPIC2015 without anonymization (0.98 with the heuristic algorithm and 0.89 with the inductive algorithm), this result can be interpreted as the impact of data distortion from anonymization being more pronounced in the sub-logs.

\mypar{RQ2: Effect of Partitioning Order on SaCoFa}
For BPIC2015 and BPIC2017, partitioning first showed slightly better performance, but overall the differences were minimal. In BPIC2012, partitioning later performed slightly better. Looking back at the earlier results with DF-Laplace, where applying partitioning last significantly reduced precision—leading to much better performance when partitioning was done beforehand—this outcome can be interpreted differently for SaCoFa. Since SaCoFa does not significantly degrade precision even when anonymization is applied without partitioning, this likely explains the observed results.

However, when applying FHM to BPIC2015, partitioning first resulted in significantly better utility compared to partitioning later. This is an unusual result, considering that for the same dataset, when using Günther’s technique or Random clustering, there was little difference between partitioning first and partitioning later. Comparing this with the results of using SaCoFa for anonymization alone (\autoref{fig:sacofa_eval35}), we see that when FHM was applied to anonymized data, precision dropped significantly. In contrast, when other techniques were applied, the precision remained relatively stable. This suggests that the effectiveness of an abstraction technique used for partitioning can vary.

\begin{figure} [htb]
    \centering
    \includegraphics[width=0.9\linewidth]{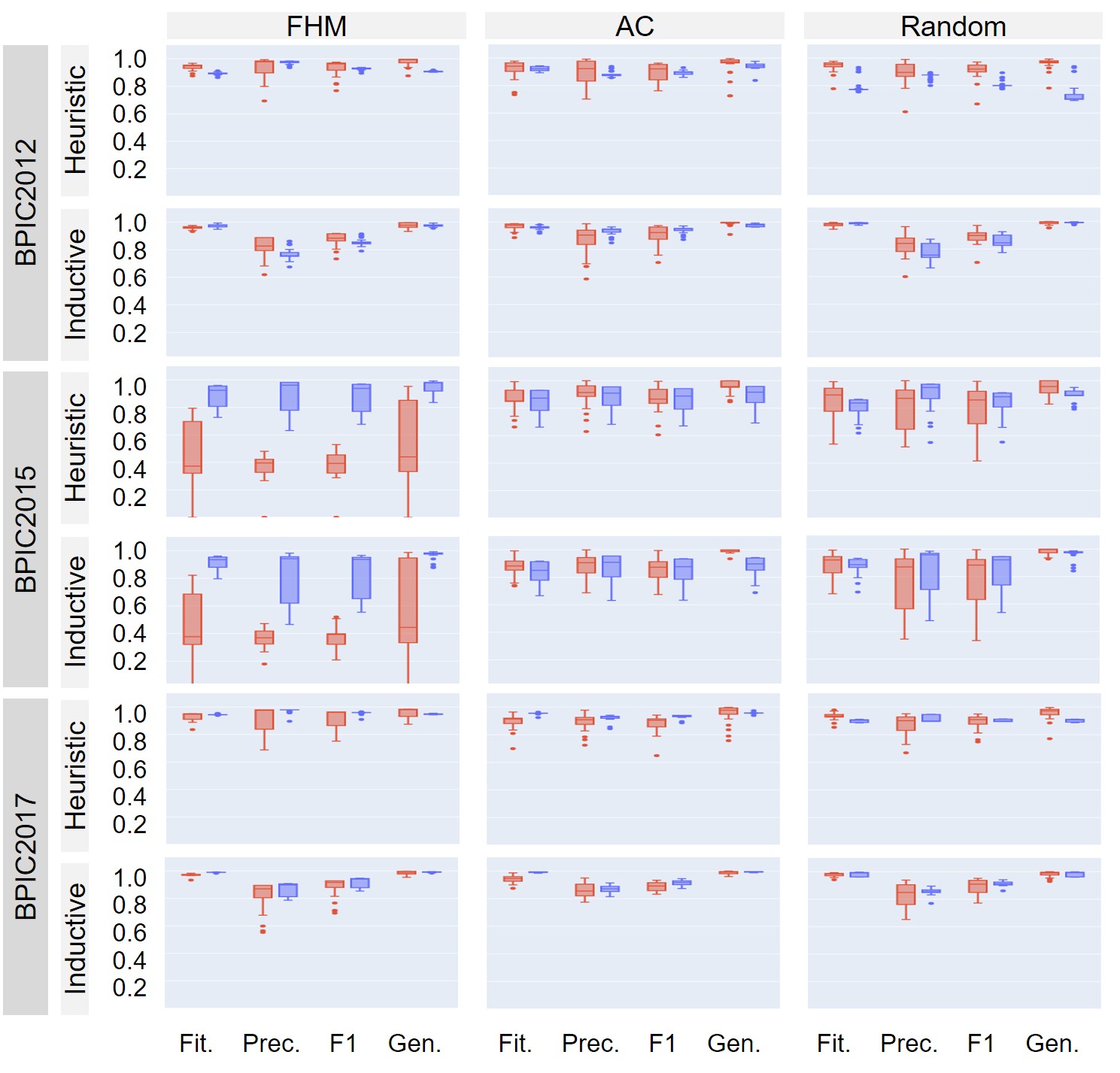}
    \caption{Utility comparison with SaCoFa: Anonymization and Partitioning (red) vs. Partitioning and Anonymization (blue).}
    \label{fig:sacofa_eval45}
\end{figure}

\subsection{Trade-Off Discussion}
\label{sec:trade_offs}
In this section, we discuss trade-offs~\cite {robillard2024communicating} that come from integrating event data partitioning into the anonymization process.

\mypar{Loss of Information due to Partitioning}
Event log partitioning can remove information that is useful for the analysis of the business process. Therefore, event partitioning can lead to an additional information loss that is difficult to quantify.
This trade-off has to be considered when weighing the utility gains from lower noise addition.

\mypar{Explosion of Pipeline Options}
Choosing the right anonymization technique is challenging as there is no standardized guide. This creates a risk that users may apply a suboptimal anonymization technique for their analytical needs.
In this paper, we introduce a pipeline that further increases the complexity of this decision. Instead of selecting only an anonymization technique, users must also choose an abstraction technique for event data partitioning. As a result, one drawback of our pipeline is that identifying the optimal configuration for maximizing utility in each scenario requires substantial domain expertise.

\mypar{Less Uncertainty of the Sub-processes}
When anonymizing the whole process, a lot of behavior can be added as noise. However, when sub-processes are anonymized independently, the new behavior that can be added is significantly restricted. This may lead an adversary to learn more information about a sub-process than would be possible using anonymization without event partitioning.
It is important to remember that differential privacy guarantees that the same privacy protection is given to individuals either way.

%% file: 5_conclusion.tex
\section{Conclusion}
\label{sec:conclusion}
In this paper, we address the problem of privacy-aware process discovery in a novel way. We propose a pipeline that builds on event data partitioning before anonymization. To evaluate our pipeline, we investigated the impact of event partitioning on anonymization for automatic process discovery.
Our findings demonstrated that our pipeline provides outperformance when directly-follows-based anonymization techniques are applied. 
We believe this findings suggest that combining anonymization with event log pre-processing  has the opportunity to unlock higher utility in privacy-aware process mining.
For future work, we intend to explore how other pre-processing techniques can contribute to better utility. 
Ideally, we aim to develop a framework that helps determine the most effective combination of pre-processing and anonymization techniques based on event log characteristics.